\documentclass[11pt,oneside,letterpaper]{article}
\usepackage{amssymb}
\usepackage{amsmath}
\usepackage{amsthm}
\usepackage[dvips]{graphicx}
\usepackage{setspace}
\usepackage{amsfonts}
\usepackage{fancyhdr}
\usepackage{xcolor}
\usepackage{graphicx}
\usepackage{subfig} 
\usepackage{rotating}
\usepackage{comment}
\usepackage{color}
\usepackage{cite}
\usepackage{braket}
\usepackage{caption}
\usepackage[abs]{overpic}
\usepackage{rotating}
\usepackage[utf8]{inputenc}
\usepackage{mathtools}

\definecolor{darkgreen}{rgb}{0,0.5,0}
\definecolor{darkblue}{rgb}{0,0,0.6}
\definecolor{purple}{rgb}{0.4,.2,0.7}

\newcommand{\be}{\begin{equation}}
\newcommand{\ee}{\end{equation}}

\usepackage[colorlinks=true,citecolor=darkgreen,linkcolor=black,urlcolor=purple]{hyperref}
\usepackage{cleveref} 

\usepackage{pdfsync}

\makeatletter
\newcommand*{\defeq}{\mathrel{\rlap{%
                     \raisebox{0.3ex}{$\m@th\cdot$}}%
                     \raisebox{-0.3ex}{$\m@th\cdot$}}%
                     =} 
\makeatother

\def\be{\begin{eqnarray}}
\def\ee{\end{eqnarray}}

\newcommand{\bea}{\begin{eqnarray}}
\newcommand{\eea}{\end{eqnarray}}
\def\ben{\begin{equation}}
\def\een{\end{equation}}

     \let\r=v

\def\be{\begin{equation}}
\def\ee{\end{equation}}
\def\ba{\begin{array}}
\def\ea{\end{array}}

\def\ba#1\ea{\begin{align}#1\end{align}}
\def\bs#1\es{\begin{split}#1\end{split}}

\allowdisplaybreaks  

\numberwithin{equation}{section}

\def \be {\begin{equation}}
\def \ee {\end{equation}}
	
\def \JM#1 {{\color{blue}  JM: #1 }}
\def \AAl#1 {{\color{red}  AA: #1 }}

\begin{document}
\onehalfspacing

\begin{center}

~
\vskip5mm

{\LARGE  {
A Traversable Wormhole from the Kerr Black Hole
\\
\ \\
}}

Ryan Bilotta

\vskip5mm
{\it Department of Physics, Cornell University, Ithaca, New York, USA
} 

\vskip5mm

{\tt rjb424@cornell.edu}

\end{center}

\vspace{4mm}

\begin{abstract}
\noindent
The approach of Gao, Jafferis, and Wall to perturbatively construct traversable wormholes has seen success in a number of black hole backgrounds, particularly BTZ and $AdS_2$, whereas historically most wormhole solutions have been either found to violate the achronal ANEC, violate a classical no-go theorem, or exist only in astrophysically irrelevant spacetimes. In this work, we show that a double-trace deformation to the near-horizon, near-extremal region of Kerr yields a traversable wormhole. We also comment on the potential for a fully nonperturbative approach to a four-dimensional rotating traversable wormhole in asymptotically flat space.

 \end{abstract}

\pagebreak
\pagestyle{plain}

\setcounter{tocdepth}{2}
{}
\vfill

\ \vspace{-2cm}
\renewcommand{\baselinestretch}{1}\small
\tableofcontents
\renewcommand{\baselinestretch}{1.15}\normalsize


\section{Introduction}\label{s:intro}


Over the last few years, wormholes have played a pivotal role in the development of our understanding of quantum gravity. Notably \cite{Hartman_2020,Bousso_2022}, significant progress has been made in understanding the information paradox through the use of Euclidean replica wormholes in the semiclassical approximation to the path integral. Nevertheless, it is both instructive and phenomenologically interesting to understand Lorentzian constructions of wormholes. Though there is a vast literature on wormhole solutions to Einstein gravity in four dimensions, up until recently most wormholes were either rendered nontraversable by a classical no-go theorem or required exotic matter in violation of the null energy condition. 

Traversable wormholes are geometries where there exists a null geodesic connecting two asymptotic boundaries. For this reason, the Raychaudhuri equation for null geodesics requires solutions to violate the average null energy condition (ANEC):
\begin{equation}
		\int_\gamma T_{\mu\nu}k^\mu k^\nu d\lambda\geq 0
\end{equation}
where the integral of the stress energy tensor is evaluated along some null geodesic $\gamma$ with tangent vector $k^\mu$ and affine parameter $\lambda$. Although the ordinary null energy condition is violated in quantum field theory, it has been proven that the ANEC holds along achronal geodesics in the absence of gravity \cite{Faulkner_2016,Hartman_2017} and is conjectured to be true in gravity \cite{Graham_2007}. Therefore traversable wormholes can be constructed to violate the ANEC so long as there exists a shorter geodesic which connects the two asymptotic regions and does not pass through the wormhole.

With these constraints in place, the authors of \cite{Gao_2017} proposed adding a relevant double trace deformation to the action between operators at the two boundaries. Working in the eternal BTZ black hole, they showed that semiclassically, even when the deformation is turned on for some finite time, the correction to the interior geometry due to backreaction from the boundary permits a null ray from one boundary to traverse through the wormhole (the averaged null energy becomes negative along this geodesic). This is a perturbative calculation which does not rely on knowing the full backreacted geometry. Although this coupling between the boundaries is acausal and non-local from a bulk observer's perspective, it can be thought of as arising from causal interactions between the mouths in some shared bulk intermediary space. The wormhole is unstable and only supports a finite amount of information before it collapses. Regardless, this is a novel example of a traversable wormhole supported by quantum effects. Subsequent work \cite{Maldacena_2017,maldacena2018eternal,Caceres_2018, Bak_2018,Ahn_2022,Al_Balushi_2021,Emparan_2021} has expanded this procedure to $AdS_2$, rotating BTZ, AdS-Schwarzschild, and multi-mouth systems.\footnote{It should be noted that other new constructions for traversable wormhole solutions in four and lower dimensions have since been introduced \cite{Horowitz_2019,Fu_2019,McBride_2019,Marolf_2019} which do not directly rely on the aforementioned methods, but rather introduce new global or topological devices which play the part of identifying the wormhole mouths. These techniques will not be explored in this paper.} This deformation has also been realized as a quantum circuit to explore the ER=EPR relation \cite{Jafferis_2022}.

The technique introduced in \cite{Gao_2017} does not provide a complete, nonperturbative description of the wormhole geometry. Incredibly, the authors of \cite{maldacena2018traversable} were able to write down an explicit four-dimensional wormhole geometry rendered traversable nonperturbatively by a negative average null energy generated by quantum fields propagating in the throat. The wormhole geometry is constructed explicitly in Einstein Maxwell theory plus charged, massless fermions. The fermions provide the negative energy required to keep the wormhole open. Specifically, for two black holes of magnetic charge $q$, a single four-dimensional chiral fermion leads to a series of degenerate Landau levels. For a large magnetic field, the states in the lowest Landau level are localized along the magnetic field lines of the black holes. The degeneracy of this state gives rise to $q$-many massless two-dimensional chiral fermions which lie along a circle going through the wormhole throat and in the exterior between the mouths. This results in a negative Casimir energy that is parametrically controlled by the number of fermions, or the charge $q$ of the black holes. This solution has a smooth interior, but it is only traversable for low energy waves. Just as in the perturbative constructions, the interior of the wormhole is unstable to large fluctuations in energy. This work was later extended to show that by tuning specific parameters in the wormhole, it could be made safe for human travel \cite{maldacena2020humanly}.

There were two attributes to this calculation that made it unique. First, it gave a nonperturbative, geometric construction for a traversable wormhole, and second, the solution was in four spacetime dimensions. If tractable solutions for traversable wormholes exist in $(3+1)$-dimensional Einstein gravity, it becomes worthwhile to ask if we may be able to see such phenomena in our own universe. Unfortunately, though their existence as primordial black holes has been considered \cite{maldacena2020comments}, magnetic black holes are still purely theoretical objects. On the other hand, rotating black holes do exist and have been observed in our universe. Recent work has studied how extremal Kerr black holes could be used as probes of new physics \cite{Horowitz_2023}.

With this in mind, the purpose of this paper is to construct a traversable wormhole in the Kerr background using the methods developed in \cite{Gao_2017}. However, in Kerr this procedure is complicated due to there being no regular, stationary vacuum state \cite{Ottewill_2000}. Notably, the Hartle-Hawking vacuum is irregular for superradiant bosonic fields. We circumvent this issue by deforming with respect to scalar fields averaged over the Kerr spheroid. These fields are not superradiant, and their coupling on the boundaries of near-extremal Kerr induces a negative null energy. Finally, we consider what a traversable Kerr wormhole may look like under the nonperturbative treatment from \cite{maldacena2018traversable}. These results may be relevant to black hole binaries in exotic entangled states.

This paper is organized as follows. In \cref{s:NHEK}, we review (near) extremal, near-horizon Kerr and its isometries. In \cref{s:Wormhole}, we give a brief description of free scalar theory in near-horizon Kerr and show that when the two timelike boundaries of Rindler near-horizon Kerr are coupled using a double trace deformation, the ANE is negative along a null ray moving from one boundary to the other through the bulk. Lastly, \cref{s:Discussion} discusses how irregularity in the state for Kerr black holes affects our description for a traversable wormhole, as well as what effects superradiance might have. We also comment on what a traversable wormhole in Kerr may look like nonperturbatively. The appendix provides a review for how superradiance in Kerr leads to irregularity in the state of the system off-axis.


\section{Review of Near-Horizon Extremal Kerr}\label{s:NHEK}


This section will serve as a brief introduction to near-horizon, near-extremal Kerr (nNHEK). In Boyer-Lindquist coordinates, the Kerr metric takes the form
\begin{equation}
\begin{gathered}
    ds^2=-\frac{\Delta}{\hat{\rho}^2}(d\hat{t}-a\sin^2\theta d\hat{\phi})^2+\frac{\sin^2\theta}{\hat{\rho}^2}((\hat{r}^2+a^2)d\hat{\phi}-ad\hat{t})^2+\frac{\hat{\rho}^2}{\Delta}d\hat{r}^2+\hat{\rho}^2d\theta^2 \\
    \Delta\equiv\hat{r}^2-2M\hat{r}+a^2,\quad
    \hat{\rho}^2\equiv\hat{r}^2+a^2\cos^2\theta,\quad a=\frac{J}{M}
\end{gathered}\label{eq:Kerr}
\end{equation}
where we take natural units $G_N=c=\hbar=1$ and assume $J>0$ for the remainder of the paper. The black hole has horizons
\begin{equation}
    r_\pm=M\pm\sqrt{M^2-a^2}
\end{equation}
corresponding to the coordinate singularities at $\Delta=0$. The Hawking temperature, horizon angular velocity, and Bekenstein-Hawking entropy follow as
\begin{align}
    T_H&=\frac{r_+-r_-}{8\pi Mr_+}\\
    \Omega_H&=\frac{a}{2Mr_+}\\
    S&=2\pi Mr_+=\frac{A}{4}
\end{align}
We can take the extremal limit by assuming the angular moment takes its maximum allowable value $J=M^2$, notably leading to $r_+=r_-=M$ and $T_H=0$. In this work, we will be in the near-extremal regime of Kerr. We can keep infinitesimal excitations above extremality by taking the Hawking temperature $T_H$ to zero while holding the ``near-horizon temperature", defined by $T_R\equiv\frac{2MT_H}{\lambda}$, fixed by sending $\lambda\rightarrow 0$. In other words, though the full Kerr metric may have zero temperature as measured from asymptotically flat infinity, an observer in the near-horizon throat will measure a finite temperature due to an infinite blueshift. Following \cite{Bardeen_1999,Bredberg_2010}, we observe that the distance to the horizon of extremal Kerr is infinite. We can then take a co-rotating, near-horizon limit
\begin{align}
		t&=\lambda\frac{\hat{t}}{2M}\\
    r&=\frac{\hat{r}-r_+}{\lambda r_+}=\frac{\hat{r}-M}{\lambda M}-\frac{2\pi T_R\hat{r}}{M}+O(\lambda)\\
		\phi&=\hat{\phi}-\Omega_H\hat{t}
\end{align}
In this near-extremal limit, our black hole parameters become
\begin{equation}
\begin{gathered}
    r_+=M+2\pi MT_R\lambda+O(\lambda^2)\\
    a=M-2\pi^2MT_R^2\lambda^2+O(\lambda^3)\\
    \sqrt{J}=M-\pi^2MT_R^2\lambda^2+O(\lambda^3)
\end{gathered}
\end{equation}
and the nNHEK metric follows as $\lambda\rightarrow 0$
\begin{equation}
\begin{gathered}
    ds^2=2J\Gamma(\theta)(-r(r+4\pi T_R)dt^2+\frac{dr^2}{r(r+4\pi T_R)}+d\theta^2+\Lambda(\theta)^2(d\phi+rdt)^2)\\
    \Gamma(\theta)=\frac{1+\cos^2\theta}{2},\quad\Lambda(\theta)=\frac{2\sin\theta}{1+\cos^2\theta}
		\label{eqn:gnNHEK}
\end{gathered}
\end{equation}
Sending $T_R\rightarrow 0$ recovers near-horizon extremal Kerr (NHEK) identically. However, whereas the Poincar\'e-like coordinates for NHEK draw analogy to $AdS$, the above coordinates for near-NHEK are most similar to thermal Rindler coordinates in $AdS$ or extended Schwarzschild coordinates for the eternal black hole. The NHEK metric describes at each value of $\theta\in\lbrack 0,\pi\rbrack$ a warped $AdS_3$ spacetime \cite{Anninos_2009} quotiented by identification under $\phi\sim\phi+2\pi$ with an enhanced local $SL(2,\mathbb{R})\times U(1)$ isometry group. Nevertheless, NHEK appears very similar to $AdS_2\times S^2$. Notably, the metric is identical to $AdS_2$ along the axis $\theta=0,\pi.$

Unlike Kerr, nNHEK is not asymptotically flat. It has two timelike boundaries which should be understood as being joined to Minkowski space to recover the full Kerr solution. This coordinate patch of nNHEK is attractive as a setting to induce a traversable wormhole because fields in the left and right Rindler wedges are causally disconnected, similar to BTZ in \cite{Gao_2017}.

\begin{figure}[!t]
	\centering
	\hspace{-0.9cm}\subfloat[]{\includegraphics[width=4.2 in, height=2.2 in]{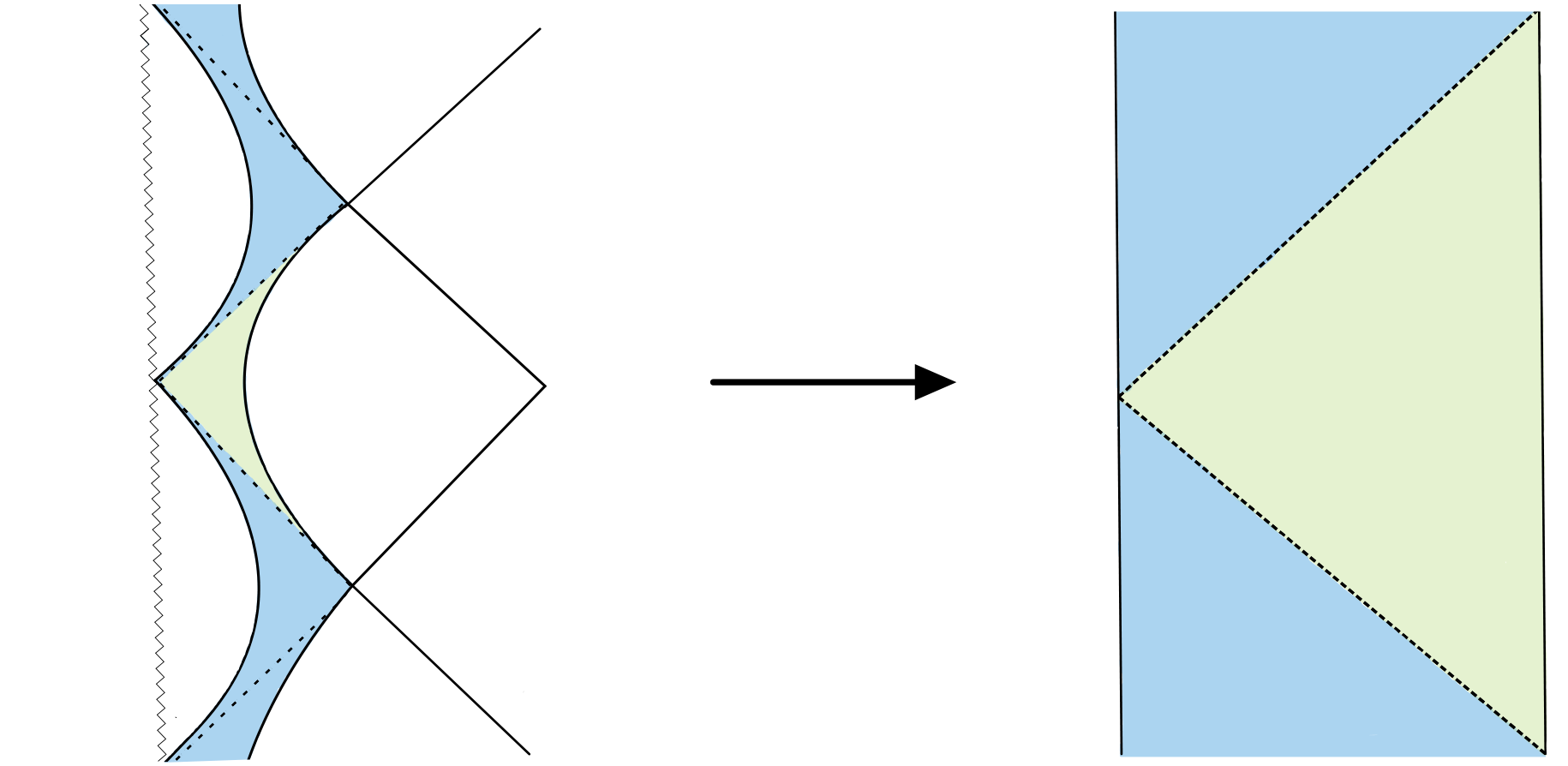}\label{fig:NHEK}}\\
	\centering\hspace{-0.45in}
\subfloat[]{\begin{overpic}[abs,unit=1mm,scale=0.25,width=4.35in, height=2.05in]{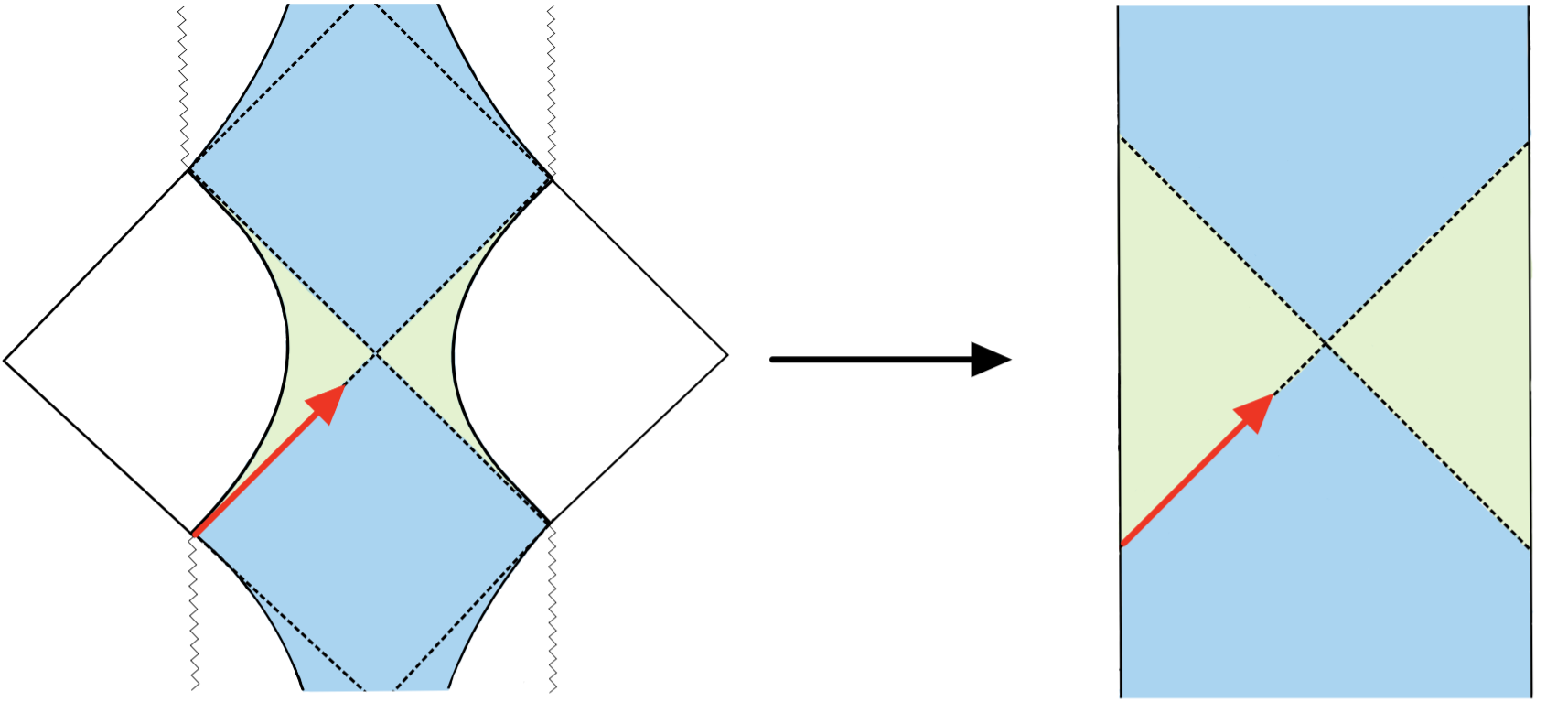}
	\centering
	\put(99.5,35){\tiny{\begin{turn}{47}
$V=0$
\end{turn}}}
	\put(81.5,41){\tiny{\begin{turn}{-48}
$U=0$
\end{turn}}}
	\put(108.5,32){\tiny{\begin{turn}{-90}
$UV=-1$
\end{turn}}}
\end{overpic}\label{fig:nNHEK}}
	\caption{Near-horizon regions described by \protect\subref{fig:NHEK} extremal Kerr and  
  \protect\subref{fig:nNHEK} near-extremal Kerr, along with their global extensions. A null ray originating on the horizon $V=0$ from the left Rindler wedge will move off the horizon and into the right Rindler wedge after backreaction from a double trace deformation couples the left and right boundaries.}
	\label{fig:nNHEKpics}
\end{figure}

The coordinates provided above for nNHEK only cover a section of the full geometry, similar to how Rindler coordinates only cover a section of $AdS$. In fact, nNHEK has a global extension given by 
\begin{equation}
    ds^2=2J\Gamma(\theta)\left(\frac{-d\tau^2+d\sigma^2}{\cos^2\sigma}+d\theta^2+\Lambda(\theta)^2(d\varphi+\tan\sigma d\tau)^2\right)
\end{equation}
where the transformation between the two coordinate systems is provided by
\begin{equation}
\begin{gathered}
    r=2\pi T_R\frac{\cos\tau-\cos\sigma}{\cos\sigma}\quad,\quad t=-\frac{1}{4\pi T_R}\ln\left|\frac{\sin\sigma-\sin\tau}{\sin\sigma+\sin\tau}\right|\\
		\phi=\varphi+\frac{1}{2}\ln\left|\frac{\sin\left(\sigma-\tau\right)}{\sin\left(\sigma+\tau\right)}\right|+2\pi T_Rt
\end{gathered}
\end{equation}
The timelike boundaries lie on $\sigma=\pm\frac{\pi}{2}$, and have an infinite proper distance between them. This is a key feature that will be exploited to create a traversable wormhole. The Poincar\'e and global coordinate systems help make the full $SL(2,\mathbb{R})$ isometry evident in the Killing symmetries generated by $\partial_t$ and $\partial_\tau$ respectively. Due to the geometry's apparent similarity to Rindler $AdS$, one might expect \cite{Guica_2009,Rasmussen_2010} that near-NHEK is dual to a nonchiral CFT. Evidence towards this has been shown more directly in \cite{Bredberg_2010,Hartman_2009}. 

Lastly, Kruskal coordinates will be important in \Cref{s:Wormhole} for parameterizing null rays moving through the wormhole. Defining the inverse temperature $\beta=\frac{1}{T_R}$, these coordinates are defined in the Rindler patch by 
\begin{equation}
\begin{gathered}
UV=-\frac{r}{r+\frac{4\pi}{\beta}}\\
-\frac{V}{U}=e^{-\frac{4\pi}{\beta}t}
\label{eq:Kruskal2}
\end{gathered}
\end{equation}
where the horizons sit at $U=0$ and $V=0$ and the boundaries lie on $UV=-1$. The metric then takes the form
\begin{equation}
ds^2=2J\Gamma\left(-\frac{4dUdV}{(1+UV)^2}+d\theta^2+\Lambda^2\left(d\phi+\frac{1}{1+UV}(UdV-VdU)\right)^2\right)
\label{eqn:gKruskal2}
\end{equation}
These coordinates have the nice property that $g_{\phi U}=g_{\phi V}=0$ on the horizon, consistent with a co-rotating reference frame.

Kruskal coordinates on NHEK are related to Kruskal coordinates on Kerr. Namely, the Kerr-Kruskal coordinates are
\begin{equation}
\begin{gathered}
\hat{U}\hat{V}=-\frac{1}{\kappa_+^2}e^{2\kappa_+r^*}\\
-\frac{\hat{V}}{\hat{U}}=e^{-2\kappa_+\hat{t}}
\label{eq:KerrKruskal}
\end{gathered}
\end{equation}
where $\kappa_+=2\pi T_H$ is the surface gravity on the outer horizon and $r^*$ is the tortoise coordinate
$$r^*=\hat{r}+\frac{r_++r_-}{r_+-r_-}\bigg(r_+\ln\left(\frac{\hat{r}-r_+}{r_+}\right)-r_-\ln\left(\frac{\hat{r}-r_-}{r_-}\right)\bigg)$$
In the near-horizon, near-extremal limit, $U=\kappa_+\hat{U}$ and $V=\kappa_+\hat{V}$.

\section{nNHEK Traversable Wormholes}\label{s:Wormhole}


A traversable wormhole induced by a double trace deformation was first introduced in \cite{Gao_2017}. Inspired by the success of \cite{maldacena2018eternal} for $AdS_2$, we will show in this section how the same procedure can be applied in nNHEK.\footnote{At infinity, this deformation \cite{Aharony_2005} has an interpretation in $AdS_{d+1}$ as a deformation to the boundary theory, where it modifies the boundary conditions on the dual bulk field data.} The deformation to the action is chosen to be 
\begin{equation}\label{eq:DTD}
		\delta S=\int dt\lambda(t)\int\ d\vec{x}_L \sqrt{\sigma_L}\Phi_L(\vec{x}_L,r_c,-t)\int d\vec{x}_R\sqrt{\sigma_R}\ \Phi_R(\vec{x}_R,r_c,t)
\end{equation}
where we have used the Rindler coordinates of nNHEK, $\vec{x}=(\theta,\phi)$, and $\Phi_L$ and $\Phi_R$ are bulk scalar operators in the left and right Rindler wedges in Kerr living on a fixed cut-off surface $r=r_c$. Notably, we integrate each field over its respective two-surface $\sqrt{\sigma}=2J\sin\theta$. The reason for individually smearing each operator over the spheroid will be explained later in this section. Lastly, the dimensionful coupling $\lambda(t)$ is used to turn the deformation on at some time $t_R=-t_L=t_0$ and switched off at some later $t_f$. We will calculate the null energy for several cases, including when the deformation is left on indefinitely and when it is switched off at a finite $t_f$.

This deformation can be thought of as arising from integrating out fields exterior to the wormhole mouths (the Rindler boundaries of NHEK) that are coupled to $\Phi_L$ and $\Phi_R$. 
\begin{equation}\label{eq:embedded}
		\delta S\sim\int dt_Ldt_R\ d\vec{x}_Ld\vec{x}_R \sqrt{\gamma_L\gamma_R}\Phi_L(\vec{x}_L,r_c,t_L)\Phi_R(\vec{x}_R,r_c,t_R)G(t_L+t_R,\vec{x}_L-\vec{x}_R)
\end{equation}
where $G$ is a propagator between $\Phi_L$ and $\Phi_R$ in the exterior space. In this sense, we use the deformation \eqref{eq:DTD} as a toy model for \eqref{eq:embedded} in NHEK. The coupling $\lambda$ is then a tensor symmetric in $\vec{x}_L$ and $\vec{x}_R$ in order to keep the deformation covariant, but for the purposes of this paper the explicit dependence is a choice of theory.

It is the backreaction from the insertion of the deformation in the bulk which provides the negative energy necessary for signals to reach the other boundary. A null ray emanating from the past horizon $\hat{V}=0$ at fixed $(\theta,\phi)$ is parameterized in Kerr-Kruskal coordinates by
\begin{equation}\label{eq:shift}
		\hat{V}(\hat{U})=-(2g_{\hat{U}\hat{V}}(\hat{V}=0))^{-1}\int_{-\infty}^{\hat{U}}d\hat{U}' h_{\hat{U}\hat{U}}(\hat{U}')
\end{equation}
where $g_{\mu\nu}$ is the Kerr metric. $\delta g_{\mu\nu}=h_{\mu\nu}$ is the variation of the background metric arising from a perturbation in the stress tensor. The condition for traversability is that $\Delta V\equiv V(\infty)<0$, where $\Delta V$ can be interpreted as the length of the wormhole's throat.

We can relate the perturbation to the ANE via the linearized Einstein equation on the horizon $\hat{V}=0$ for Kerr. By choosing an appropriate ansatz $h_{\mu\nu}(\hat{U},\theta,\phi)=(1+\frac{r_-}{r_+}\cos^2\theta)f_{\mu\nu}(\hat{U},\theta,\phi)$, we can find that to first order in $\kappa_+$,
\begin{align}
8\pi T_{\hat{U}\hat{U}}&=\frac{\kappa_+}{r_+}\Big[\left(2f_{\hat{U}\hat{U}}+\hat{U}\partial_{\hat{U}}f_{\hat{U}\hat{U}}\right)-\Gamma(\theta)\left(\left(\partial_\phi+\partial^2_\phi\right)f_{\hat{U}\hat{U}}-2\partial_{\hat{U}}\partial_\phi f_{\phi \hat{U}}+\partial^2_{\hat{U}}f_{\phi\phi}\right)\Big]\notag\\
&-\frac{1}{2r_+^2}\Big[\big(\partial_\phi+\frac{1}{\Lambda(\theta)^2}\partial^2_\phi+\left(\cot\theta+\partial_\theta\right)\partial_\theta\big)f_{\hat{U}\hat{U}}-2\left(\cot\theta+\partial_\theta\right)\partial_{\hat{U}}f_{\theta \hat{U}}\notag\\
&\hspace{1.5cm}+\partial^2_{\hat{U}}f_{\theta\theta}-\frac{1}{\Lambda(\theta)^2}\left(2\partial_{\hat{U}}\partial_\phi f_{\phi \hat{U}}-\partial^2_{\hat{U}}f_{\phi\phi}\right)\Big]
\end{align}
Upon integrating over $\hat{U}$, total derivatives in $\hat{U}$ drop out due to the asymptotically flat boundary conditions on Kerr. 
\begin{equation}
8\pi\int d\hat{U}T_{\hat{U}\hat{U}}=\Big[\frac{\kappa_+}{r_+}\big(1-\Gamma(\theta)\left(\partial_\phi+\partial^2_\phi\right)\big)-\frac{1}{2r_+^2}\big(\partial_\phi+\frac{1}{\Lambda(\theta)^2}\partial^2_\phi+\left(\cot\theta+\partial_\theta\right)\partial_\theta\big)\Big]\int d\hat{U} f_{\hat{U}\hat{U}}
\end{equation}
\indent At this stage, we could proceed as in \cite{Marolf_2019,Fu_2019} with the task of finding the appropriate Green's function $H(\Omega,\Omega')$ to invert this equation
\begin{equation}
\left(\int d\hat{U} f_{\hat{U}\hat{U}}\right)\left(\Omega\right)=8\pi\int d\Omega'H(\Omega,\Omega')\int d\hat{U}T_{\hat{U}\hat{U}}(\Omega')
\end{equation}
However, in our case, we are just interested in the question of traversability and do not need the ANE's full angular dependence. It is thus sufficient to integrate over the unit spheroid $d\Omega=\sin\theta d\theta d\phi$:
\begin{equation}
8\pi \int d\Omega d\hat{U}T_{\hat{U}\hat{U}}=\frac{\kappa_+}{r_+}\int d\Omega d\hat{U} \frac{h_{\hat{U}\hat{U}}}{1+\cos^2\theta}
\label{eq:LEE}
\end{equation}
The right-hand side of \eqref{eq:LEE} is measuring $\Delta \hat{V}_{\text{avg}}$, a shift in the null ray averaged over the sphere \cite{Fu_2019}. Combining Eqs. \eqref{eq:LEE} and \eqref{eq:shift}
\begin{equation}
T_H^3\Delta \hat{V}_{\text{avg}}=\frac{1}{4\pi^2 M}\int d\Omega d\hat{U}T_{\hat{U}\hat{U}}
\end{equation}
In the near-horizon, near-extremal limit, the averaged stress energy reduces to the nNHEK case
\begin{equation}
T_H^2\Delta \hat{V}_{\text{avg}}=\frac{1}{2\pi M}\int d\Omega dU T_{UU}
\label{eq:DeltaV}
\end{equation}
Hence, the stress energy's average value over the spheroid is sufficient to determining traversability. This is a motivating factor for considering the deformation \eqref{eq:DTD}, which is a kind of double trace deformation of fields averaged over the spheroid. However, it appears that in the extremal limit $T_H\rightarrow0$, $\hat{V}(\infty)$ in \eqref{eq:shift} diverges, corresponding to a diverging zero mode in $H(\Omega,\Omega')$. This feature is also present in \cite{Marolf_2019,Fu_2019,Caceres_2018}, and is likely attributed to the Aretakis instability for gravitational perturbations on extreme Kerr \cite{Strominger_1999,Lucietti_2012}.\footnote{The instability for scalar perturbations on Kerr \cite{Aretakis_2013} is also related to the issue of vacuum regularity, discussed in \Cref{app:Regularity}.} This indicates a breakdown in the first order perturbation above, while indicating that a non-perturbative analysis may be possible for a finite ANE as in \cite{Fu_2019,maldacena2018traversable}. Therefore, in determining the traversability of a nNHEK wormhole, we will be assuming that the Hawking temperature is small, but nonzero.

Using \eqref{eqn:gKruskal2}, \eqref{eq:shift}, and \eqref{eq:LEE}, we see that a negative ANE will render the wormhole traversable. Note that the negative null energy will not violate the achronal ANEC. Though null rays can propagate through the bulk with negative average null energy, information can be exchanged through the boundaries via the coupling \eqref{eq:DTD}.

The stress tensor for a scalar field in a curved background is given by
\begin{equation}
		T_{\mu\nu}=\partial_\mu\Phi\partial_\nu\Phi-\frac{1}{2}g_{\mu\nu}g^{\rho\sigma}\partial_ \rho\Phi\partial_\sigma\Phi-\frac{1}{2}g_{\mu\nu}\mu^2\Phi^2
\end{equation}
This allows us to determine its 1-loop expectation value via point splitting
\begin{equation}
\left<T_{\mu\nu}\right>=\lim_{x\rightarrow x'}\left(\partial_\mu\partial'_\nu G(x,x')-\frac{1}{2}g_{\mu\nu}g^{\rho\sigma}\partial_\rho\partial'_\sigma G(x,x')-\frac{1}{2}g_{\mu\nu}\mu^2G(x,x')\right)
		\label{eqn:T}
\end{equation}
where $G(x,x')$ is the contribution to the bulk two-point function due to the deformation \eqref{eq:DTD}. The two-point function is calculated by evolving the fields in time by $U(t,t_0)=\mathcal{T}e^{-i\int_{t_0}^tdt\delta H(t)}$, where the deformation to the Hamiltonian is
\begin{equation}
\delta H(t)=-\int d\vec{x}_Ld\vec{x}_R\lambda(t)\sqrt{\sigma_L\sigma_R}\Phi_R(\vec{x}_R,r_c,t)\Phi_L(\vec{x}_L,r_c,-t)
\end{equation}
To first order in $\lambda$, the first order correction to the two-point function between fields in the right wedge is\cite{Gao_2017}
\begin{align}\label{eq:Gh}
G_\lambda(x,x')&=-i\int dt_1d\vec{x}_Ld\vec{x}_R\lambda(t_1)\sqrt{\sigma_L\sigma_R}\left<\right[\Phi_R(\vec{x}_R,r_c,t_1)\Phi_L(\vec{x}_L,r_c,-t_1),\Phi_R(x)\left]\Phi_R(x')\right>\notag\\
&\hspace{1.5in}+(x\leftrightarrow x')\notag\\
&=-i\int dt_1d\vec{x}_Ld\vec{x}_R\lambda(t_1)\sqrt{\sigma_L\sigma_R}G_{LR}(\vec{x}_L,r_c,-t_1,x')\left<[\Phi_R(\vec{x}_R,r_c,t_1),\Phi_R(x)]\right>\notag\\
&\hspace{1.5in}+(x\leftrightarrow x')
\end{align}
where $x=(t,r,\theta,\phi)$ in the Schwarzschild coordinates of nNHEK, and 
\begin{equation}
G_{LR}(\vec{x}_L,r_c,-t_1,x')=\left<\Phi_L\left(\vec{x}_L,r_c,-t_1\right)\Phi_R(x')\right>
\end{equation} 
is the bulk Wightman function in the unperturbed nNHEK spacetime. The expectation value of the commutator above is related to the retarded correlation function $G_{R}^r(x,x')=i\left<[\Phi_R(x),\Phi_R(x')]\right>\theta(t-t')$.

At this stage, the procedure would be to calculate $\left<T_{UU}\right>$ to first order in $\lambda$. However, in writing down the two-point functions above, we implicitly selected a vacuum state. A natural choice would be to work in the Hartle-Hawking state, where the state is regular on the past and future horizons and the symmetries of the state cause the expectation value of $\int dUT_{UU}$ to vanish on the horizon. The problem of constructing regular vacuum states on Kerr has long been a subject of study. Kay and Wald \cite{Wald_1988} proved that there does not exist a regular Hadamard state for scalars on Kerr which also possesses the correct symmetries of the geometry. In other words, we cannot construct a ``Hartle-Hawking"-like state in nNHEK. Candelas, Chrzanowski, and Howard \cite{Candelas_1981} and Frolov and Thorne \cite{Frolov_1989} constructed various vacuum states which possess some of the desired properties of a Hartle-Hawking state, but it has since been shown \cite{Ottewill_2000} that these states are either only regular on the axis of rotation (where the space is effectively $AdS_2$) or are not equilibrium states with the desired symmetries we should expect of a Hartle-Hawking state. This irregularity stems from two effects: classical bosonic superradiance in Kerr and the inherent irregularity of thermal states beyond the speed of light surface, the region where an observer co-rotating with the black hole horizon would need a velocity greater than the speed of light. The irregularity of the scalar vacuum in Kerr is reviewed in detail in \Cref{app:Regularity}. See also \cite{Winstanley_2023} for a nice review. On the contrary, fermions, which do not possess classical superradiance, do not run into one of the regularity problems we see with bosons, and seemingly regular vacuum states defined out to the speed-of-light surface have been constructed in \cite{Winstanley_2013,Gerard_2021}. An interesting future avenue would be to see how the results of this work differ in the case of free fermions on nNHEK.

For NHEK, this means we cannot analyze off-axis scalar mode contributions to the integrated null energy at thermal equilibrium. However, by working in the global vacuum state of nNHEK (described in the next section) and integrating over the spatial boundary in \eqref{eq:DTD}, we remove superradiant modes and are left with a regular two-point function.


\subsection{Free Scalars in NHEK}\label{ss:Scalars}


The focus of this section is to determine the bulk-to-boundary propagator in near-NHEK via a mode sum. In previous similar works on traversable wormholes, the geometries had known solutions for the two point functions, allowing them to immediately determine, analytically or numerically, the average null energy. The story in NHEK is more complicated. Briefly, the mode sum solution for the bulk two-point function is not known in Kerr. Though work has been done \cite{Dias_2009} to understand the scalar mode solutions and the spectrum in NHEK, there is no closed form for the correlation functions.

We need to solve the scalar wave equation in the near-NHEK background. Evaluating the equation in the global coordinates,
\begin{equation}\label{eq:Phi}
    \Phi\left(\tau,\sigma,\theta,\varphi\right)=\mathcal{R}\left(\sigma\right)\mathcal{S}\left(\theta\right)e^{im\varphi}e^{-i\omega \tau}
\end{equation}
where we assumed that the field is separable in time and axial angle. We can separate the scalar wave equation into radial and angular equations.
\begin{equation}
\begin{gathered}\label{eq:diff}
\frac{1}{\sin{\theta}}\frac{d}{d\theta}\left(\sin{\theta}\frac{d}{d\theta}\mathcal{S}\right)+\left(K_{lm}-(\frac{m^2}{4}-J\mu^2)\sin^2\theta-\frac{m^2}{\sin^2\theta}\right)\mathcal{S}=0\\
\frac{d}{d\rho}\left((1+\rho^2)\frac{d}{d\rho}\mathcal{R}\right)+\left(\frac{(\omega+m\rho)^2}{1+\rho^2}+m^2-K_{lm}-2J\mu^2\right)\mathcal{R}=0
\end{gathered}
\end{equation}
where $\rho=\tan\sigma$ and we have introduced a separation constant $K_{lm}$ which must be computed numerically. The angular equation is solved by the set of the spheroidal harmonics $\mathcal{S}_{lm}(\theta)$, while the radial equation is solved by a sum of hypergeometric functions with complex arguments \cite{Amsel_2009}. As explained earlier, our deformation is designed such that superradiant modes do not contribute. Namely, integrating the field over the transverse directions lead to solutions of \eqref{eq:diff} with $m=0$. The radial solution for the scalar modes simplifies to the $AdS_2$ answer \cite{NAKATSU_1999,Spradlin_1999}
\begin{equation}
\mathcal{R}_{nl}\left(\sigma\right)=\Gamma(h_l)2^{h_l-1}\sqrt{\frac{n!}{\pi\Gamma(n+2h_l)}}\cos^{h_l}(\sigma)C_n^{h_l}(\sin\sigma)
\end{equation}
where $h_l\equiv\frac{1}{2}\pm\sqrt{\frac{1}{4}+K_l+2J\mu^2}$ and $C_n^{h_l}(\sin\sigma)$ is the Gegenbauer function. We have fixed normalizable boundary conditions to discretize the spectrum $\omega_n=n+h_l$ for $n=0,1,\ldots$ and have normalized the modes with respect to the Klein-Gordon norm. The bulk two-point function is then
\begin{equation}\label{eq:bulkG}
    \left<\Phi(x)\Phi(x')\right>=\sum_{n,l,m}\Phi^*_{nlm}(x')\Phi_{nlm}(x)=\sum_{n,l,m}e^{-i\omega_n(\tau-\tau')}e^{im(\varphi-\varphi')}\mathcal{S}_{lm}(\theta)\mathcal{S}^*_{lm}(\theta')\mathcal{R}_{nlm}(\sigma)\mathcal{R}_{nlm}^*(\sigma')\notag
\end{equation}
By incorporating the boundary integral in \eqref{eq:Gh}, the Wightman functions takes the form
\begin{equation}\label{eq:Wightman}
\int d\phi\left<\Phi(x)\Phi(x')\right>=\sum_l\frac{\Gamma(h_l)^2}{2\Gamma(2h_l)}\mathcal{S}_{l}(\theta)\mathcal{S}^*_{l}(\theta')\left(\frac{2}{d_{glob}}\right)^{h_l}\ _2F_1(h_l,h_l,2h_l,-\frac{2}{d_{glob}})
\end{equation}
where we have used an identity on Gegenbauer functions \cite{Spradlin_1999} in order to perform the sum in $n$. $d_{glob}$ is the usual $SL(2,\mathbb{R})$ invariant distance function on $AdS_2$
\begin{equation}
d_{glob}\equiv\frac{\cos(\tau-\tau')-\cos(\sigma-\sigma')}{\cos(\sigma)\cos(\sigma')}
\end{equation}
In defining the two-point function above with respect to the global modes, we have selected a ``global vacuum state" analogous to a Hartle-Hawking state or the Frolov-Thorne state in Kerr. \Cref{app:Regularity} describes how this state is irregular unless it is evaluated on-axis due to superradiant mode contributions. Specifically, the two-point function is only regular when at least one of the fields is on-axis. But we also need to discuss the boundary conditions imposed on the fields. Just as in $AdS$, for sufficiently negative values of $\mu^2$, we can quantize with respect to either the positive or negative root of $h_l$. However, for NHEK, if we choose to quantize with respect to the negative root, the radial modes fail to be Klein-Gordon normalizable for $l>0$. Namely, normalizability restricts $h_l$ to be real and greater than $-\frac{1}{2}$. So, moving forward we will take $h_l$ to have the positive root. 

Finally we can determine the Wightman function $G_{LR}$ by taking the limit where one of the fields moves to the cutoff surface $r_c=\frac{2\pi}{\beta}\left(\frac{\cos(\tau)}{\cos(\sigma_c)}\right)$. We have that
\begin{align}\label{eq:GLR}
\int d\phi\ G_{LR}(\vec{x},\sigma_c,\tau,x')\approx\frac{2^{h_0}\Gamma(h_0)^2}{2\Gamma(2h_0)}\left(\frac{\cos(\sigma')\cos(\sigma_c)}{\cos(\tau'-\tau)+\sin(\sigma')}\right)^{h_0}\mathcal{S}_{0}(\theta)\mathcal{S}_{0}(\theta')
\end{align}
Notably, the $l=0$ mode is the dominant term in the mode sum for the bulk two-point function close to the boundary.

One might wonder whether or not there exists a dual description to this gravitational perturbation. Due to the nature of our averaging procedure, the averaged correlation functions are described by an $AdS_2$ solution. This is similar to the eternal traversable wormhole in nearly-$AdS_2$ in \cite{maldacena2018eternal}. From this perspective, our deformation \eqref{eq:DTD} constitutes a marginal interaction on the boundary for $h_0=\frac{1}{2}$ and irrelevant otherwise. It would be interesting in the future to understand whether or not \eqref{eq:DTD} can be modeled by coupled SYK theories at low energy as in \cite{maldacena2018eternal}.



\subsection{Negative Average Null Energy}

\begin{figure}[!htbp]
	\centering
		\includegraphics[width=4.5 in,height=2.5 in]{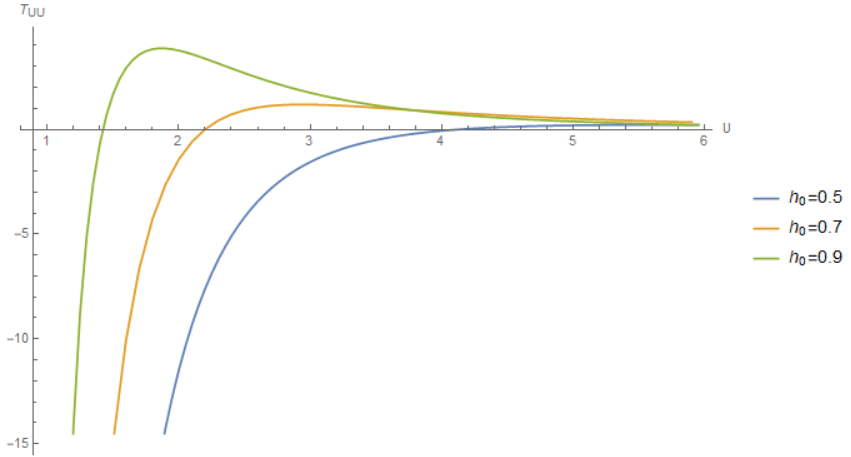}
	\caption{The null energy at $\theta=0$ as a function of $U$ for various values of $h_0$, where we turn on the deformation at $U_0=1$.\label{abc}}
\end{figure}

We wish to prove that the null energy integrated over a null geodesic along the horizon is negative. Signals moving along the future horizon receive a time advance and reach the timelike boundary of the right Rindler patch. Therefore, we will work in the Kruskal coordinates defined earlier and set $V=0$. The deformation will only have support after some turn-on time $U_0=e^{\frac{2\pi}{\beta}t_0}$
\begin{equation}
\lambda(t)=
		\begin{cases}
		\lambda_0\ r_c^{2h_0} & \text{$t\geq t_0$}\\
		0 & \text{else}
		\end{cases}
\end{equation}
where $\lambda_0$ has units of energy squared. With \eqref{eq:Wightman}, \eqref{eq:GLR}, and the transformation \eqref{eq:Kruskal2}, the first order correction to the two-point function \eqref{eq:Gh} in Kruskal coordinates is
\begin{equation}
\begin{gathered}
G_\lambda(U,\theta,U',\theta')= A \int_{U_0}^U\frac{dU_1}{U_1}\left(\frac{1}{1+U'U_1}\right)^{h_0}\left(\frac{U_1}{U-U_1}\right)^{h_0}\mathcal{S}_{0}(\theta)\mathcal{S}_{0}(\theta')+(U\leftrightarrow U')\\
\label{eq:GlambdaKruskal}
A\equiv 2\lambda_0C^2J^2\left(\frac{2^{h_0}\Gamma(h_0)^2}{\Gamma(2h_0)}\right)^2\sin(\pi h_0)\left(\frac{2\pi}{\beta}\right)^{2h_0-1}
\end{gathered}
\end{equation}
where the constant $C=\int d\theta \sin\theta\mathcal{S}_{0}(\theta)$. This constant must be computed numerically for specific values of $\mu$ and $J$.
The choice of coordinates is important here. We are evaluating the deformation on $r_c=\frac{2\pi}{\beta}\left(\frac{\cos(\tau)}{\cos(\sigma_c)}\right)$ and taking $\sigma_c$ to be very close to the boundaries $\sigma=\pm\frac{\pi}{2}$. The fixed $r_c$ surface has dependence on global time $\tau$, so solely regulating the fall-off in $\sigma$ would not yield the proper boundary conditions. Inserting \eqref{eq:GlambdaKruskal} into \eqref{eqn:T}, the null energy is
\begin{equation}\label{eq:TpreInt}
T_{UU}(U,\theta)=2A\ \mathcal{S}_{0}(\theta)^2\lim_{U'\rightarrow U}\partial_U\partial_{U'}\int_{U_0}^U\frac{dU_1}{U_1}\left(\frac{U_1}{(1+U'U_1)(U-U_1)}\right)^{h_0}\\ \end{equation}
The integral above can be performed analytically to yield 
\begin{align}\label{eq:TUU}
T_{UU}(U,\theta)=-2h_0^2A\ \mathcal{S}_{0}(\theta)^2&\Bigg(\frac{1}{U^{2+h_0}}\Big[\frac{U}{1+h_0}F_1\big(1+h_0;h_0,1+h_0;2+h_0;\frac{1}{U},-U\big)\notag\\
&\quad\quad\quad+\frac{1}{2+h_0}F_1\big(2+h_0;1+h_0,1+h_0;3+h_0;\frac{1}{U},-U\big)\Big]\notag\\
&-\pi\csc(h_0\pi)\Big[h_0\ _2F_1\left(1+h_0,1+h_0;2;-U^2\right)\\
&\quad\quad\quad-(1+h_0)\ _2F_1\left(1+h_0,2+h_0;2;-U^2\right)\Big]
\Bigg)\notag
\end{align}
The behavior of $T_{UU}$ along the axis $\theta=0$ is given in \Cref{abc}.
Although \eqref{eq:TUU} above does not seem like it would be amenable to integration analytically, we can use the same trick employed in the appendix of \cite{Gao_2017} in order to express the integral of the null energy in a relatively simple form
\begin{align}
\int_{U_0}^\infty dU T_{UU}=-A\ \mathcal{S}_{0}(\theta)^2&\frac{\Gamma(1-h_0)\Gamma(\frac{1}{2}+h_0)^2}{\Gamma(h_0)}\Bigg(1-\frac{4^{h_0}\Gamma(1+h_0)}{\sqrt{\pi}\Gamma(\frac{3}{2}+h_0)}U_0\times\\
&\times\left(\frac{U_0}{1+U_0^2}\right)^{h_0}\ _2F_1\big(1,\frac{1}{2}-h_0;\frac{3}{2}+h_0;-U_0^2\big)\Bigg)\notag
\end{align}
Notably, if we turn on the deformation at $t=0$, the averaged null energy is
\begin{align}\label{eq:intT}
\int_{U_0=1}^\infty dU T_{UU}&=-A\ \mathcal{S}_{0}(\theta)^2\frac{\Gamma(1-h_0)\Gamma(\frac{1}{2}+h_0)^2}{2\Gamma(h_0)}\notag\\
&=-4^{1-h_0}\pi^2C^2\lambda_0J^2\left(\frac{2\pi}{\beta}\right)^{2h_0-1}\mathcal{S}_{0}(\theta)^2
\end{align}
\begin{figure}[!htbp]
	\centering
		\includegraphics[width=4.5 in,height=2.5 in]{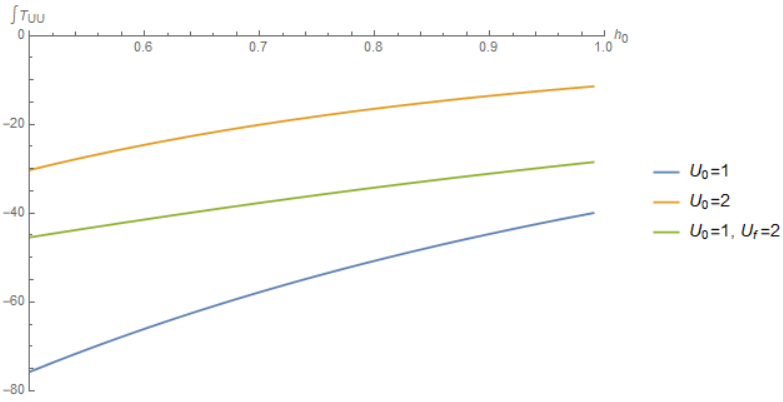}
	\label{fig:Graph}
	\caption{The integrated null energy at $\theta=0$ as a function of $h_0$. The red and blue curves correspond to turning on the deformation indefinitely at $U_0=1$ and $U_0=2$ respectively, while in the green curve the deformation is turned off at $U_f=2$.}
\end{figure}
This null energy is negative for the range $h_0\in[\frac{1}{2},1]$. The marginal value of $h_0=\frac{1}{2}$ sits at a kind of Breitenlohner-Freedman bound for scalars in nNHEK. In terms of the scalar mass and the black hole spin, $J\mu^2>-0.187$. Our result for the null energy is similar to what one would receive for a double trace deformation to pure $AdS_2$. Note that as the near-horizon temperature $T_R$ scales to zero and we move to a NHEK picture, so does the energy. This coincides with the model breaking down, as taking the extremal limit sends Rindler nNHEK to Poincare NHEK.

This wormhole is unstable. The throat will close if we throw in enough energy to cause the overall null energy along the horizon to be positive. Bounds on information transfer for traversable wormholes in $AdS_2$ have been studied in \cite{Maldacena_2017}.


\section{Discussion}\label{s:Discussion}

A double trace deformation renders the Rindler patch of nNHEK traversable through the bulk. Aside from the deformation not being relevant, there is another aspect to the calculation which may potentially weaken this conclusion: we have restricted to on-axis, non-superradiant modes in the two-point function. It may be the case that superradiant mode contributions increase the negative energy. Amplification in the superradiant sector within the wormhole could lead to a Casimir energy that grows in time. Unfortunately, as discussed earlier and as detailed in \Cref{app:Regularity}, it is not easy to determine a symmetric, regular vacuum state in Kerr off-axis. Superradiant modes yield nonnormalizable fields and scalar two-points functions which are not regular off-axis. Any calculation involving superradiant modes would thus require knowledge of the state out of thermal equilibrium both inside and outside the wormhole off-axis.

A natural question to ask from here is whether or not this deformation can be thought of as arising from an effective description in Kerr. Specifically, the question is whether it is possible to give a nonperturbative description of the traversable Kerr wormhole geometry similar to what is done with magnetic black holes in \cite{maldacena2018traversable,maldacena2020humanly}. The authors of \cite{maldacena2018traversable} took advantage of the Landau levels for a fermion in a magnetic field to generate a parametrically large negative energy to maintain their wormhole. For a rotating system, classical superradiance is a potential candidate to generate negative energy in the form of a Casimir-like effect across the wormhole. However, we once again run into the issue of vacuum regularity. It is important to know the exact state of the system in order to study potential Casimir effects generated by superradiance inside the throat and in the shared region exterior to the wormhole. This issue could be circumvented by using fermions. Bosonic superradiance can be thought of as stemming from the classical area theorem, but fermions get around this because they violate the null energy theorem on the horizon. In fact, a Hartle-Hawking state has been proposed \cite{Winstanley_2013} for fermions that is regular off-axis out to the speed-of-light surface. Despite this advantage, a lack of superradiance would deprive the system of a means to generate a large enough negative energy to sustain the wormhole. 

Assuming one could determine the out-of-equilibrium state off-axis and calculate this energy, the geometry of the wormhole would mirror that of the four-dimensional construction in \cite{maldacena2018traversable}. The global extension of nNHEK would be used to model the ``bridge" of the wormhole, while the exterior of the wormhole is provided by ordinary Kerr and its near-horizon expansion.\footnote{This assumes the mouths are sufficiently far away so that the space around them can be approximated as Kerr} Finally, the astrophysical significance of considering a traversable wormhole between rotating black holes naturally leads to wondering how such objects could affect observations of emissions from rotating binary systems. In particular, work has been done \cite{Gralla_2018,Kapec_2019,Lupsasca_2020,Hadar_2022} to predict how the symmetries of NHEK will affect measurements of high-spin black holes by the Event Horizon Telescope. It would also be interesting to consider what affects a wormhole may have on gravitational wave measurements of black hole mergers.
\begin{figure}[bhtp]
	\centering
		\includegraphics[width=4.5 in,height=1.5 in]{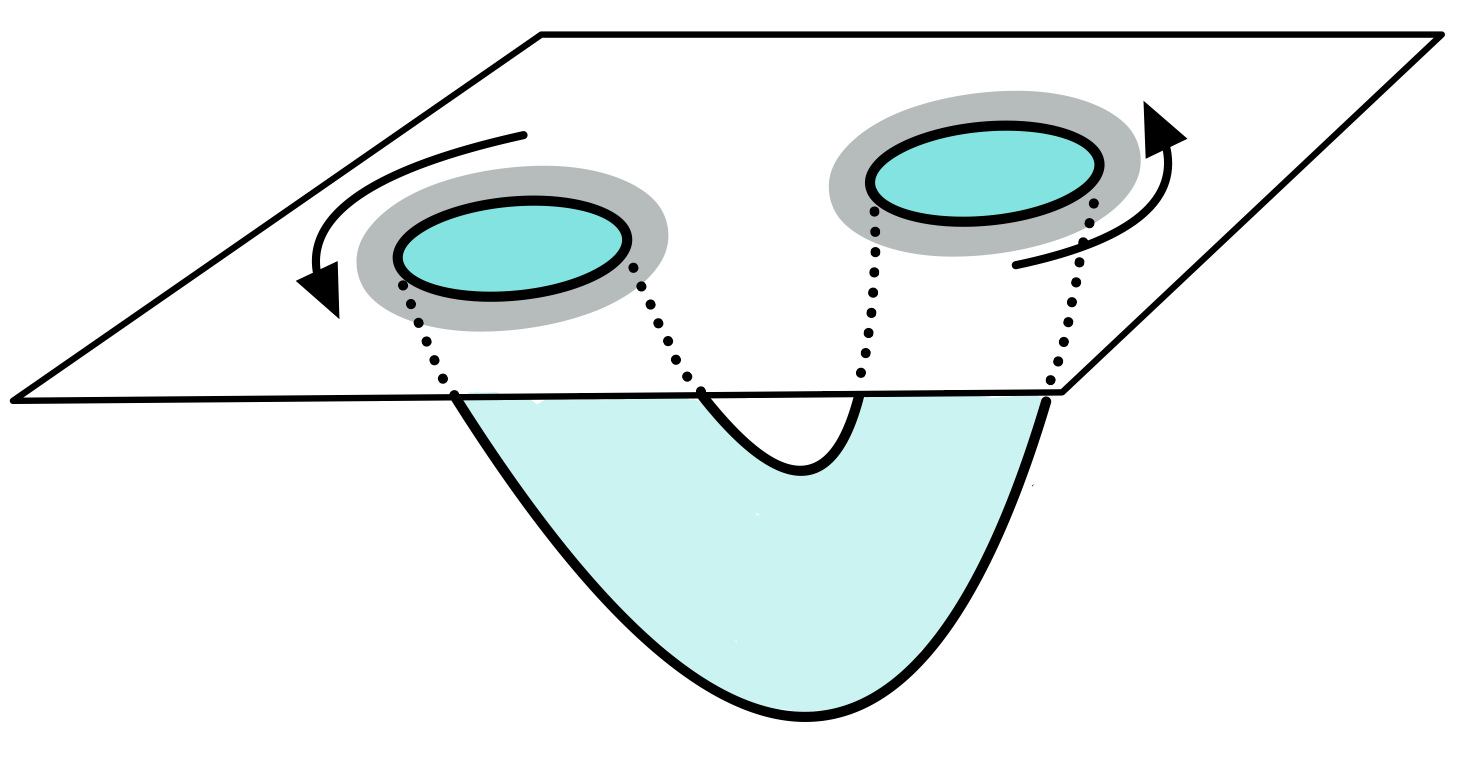}
	\caption{A particular time slicing of the four-dimensional Kerr wormhole. The wormhole itself (blue) is a global near-horizon Kerr. It connects to the asymptotically flat spacetime via a throat region (gray) described by ordinary Kerr. Just like Kerr, the throats of the wormhole should rotate. To prevent the throats from pulling together, we could add orbital rotation \cite{maldacena2018traversable}.}
	\label{fig:Wormhole}
\end{figure}


\ \\
\bigskip

\noindent \textbf{Acknowledgments}
We would like to thank Tom Hartman, Edgar Shaghoulian, Amir Tajdini, and Kanato Goto for useful discussions. This work is supported by NSF grant PHY-2014071.

\appendix

\section{Vacuum Regularity in Kerr}\label{app:Regularity}

As indicated in \Cref{s:Wormhole}, the Frolov-Thorne state in Kerr is not regular off-axis due to superradiance. This appendix will serve to show the origin of this issue. In particular, we will analyze the properties of various vacuum states in Kerr by studying their scalar two-point functions.

The mode solutions to the scalar wave equation in Kerr are separable and take the form\footnote{Note that the Boyer-Lindquist coordinates no longer have hats as compared to \eqref{eq:Kerr}. This is to ease notation in this section only.}
\begin{equation}
\Phi_{nlm}(x)=N_{\omega lm}e^{-i\omega t}e^{im\phi}\mathcal{S}_{\omega lm}(\theta)\frac{\mathcal{R}_{\omega lm}(r)}{\sqrt{r^2+a^2}}
\end{equation}
where $N_{\omega lm}$ is a normalization constant determined by the usual Klein-Gordon inner product
\begin{equation}
(\Phi_1,\Phi_2)=\frac{i}{2}\int_\Sigma d\Sigma^\mu \left((\partial_\mu\Phi^*_2)\Phi_1-\Phi^*_2(\partial_\mu\Phi_1)\right)
\end{equation}
which is independent of the choice in Cauchy surface $\Sigma$. As in NHEK, $\mathcal{S}_{\omega lm}(\theta)$ are the set of spheroidal harmonics, while the radial functions $\mathcal{R}_{\omega lm}(r)$ are solved via a Schr\"odinger-like equation
\begin{equation}\label{eq:Krad}
\left(\frac{d^2}{dr_*^2}-V_{\omega lm}(r)\right)\mathcal{R}_{\omega lm}(r)=0
\end{equation}
\begin{equation}
V_{\omega lm}(r)\sim
\begin{cases}
-\tilde{\omega}^2,&r_*\rightarrow-\infty\\
-\omega^2,&r_*\rightarrow\infty
\end{cases}
\end{equation}
where $r_*$ is the tortoise coordinate $dr_*/dr=(r^2+a^2)/\Delta$ and $\tilde{\omega}=\omega-m\Omega$. As we go to infinity, the radial functions will fall into two general classes asymptotically
\begin{align}
&\mathcal{R}^-_{\omega lm}(r)\sim
\begin{cases}\label{eq:asym}
e^{i\tilde{\omega}r_*}+A^-_{\omega l m}e^{-i\tilde{\omega}r_*},&r_*\rightarrow-\infty\\
B^-_{\omega l m}e^{i\tilde{\omega}r_*},&r_*\rightarrow\infty
\end{cases}\\
&\mathcal{R}^+_{\omega lm}(r)\sim
\begin{cases}\label{eq:asyp}
B^+_{\omega l m}e^{-i\tilde{\omega}r_*},&r_*\rightarrow-\infty\\
e^{-i\tilde{\omega}r_*}+A^+_{\omega l m}e^{i\tilde{\omega}r_*},&r_*\rightarrow\infty
\end{cases}
\end{align}
The constants above represent transmission and reflection coefficients for scalar waves that scatter off of the black hole. By way of the Wronskian relation for the radial equation, DeWitt showed that \cite{DeWitt_1975}
\begin{align}\label{eq:Kasym}
1-|A^+_{\omega lm}|^2&=\frac{\tilde{\omega}}{\omega}|B^+_{\omega lm}|^2\\
1-|A^-_{\omega lm}|^2&=\frac{\omega}{\tilde{\omega}}|B^-_{\omega lm}|^2\\
\omega B^{-\ \ast}_{\omega lm}A^+_{\omega lm}&=-\tilde{\omega}B^+_{\omega lm}A^{-\ \ast}_{\omega lm}\\
\omega B^-_{\omega lm}&=\tilde{\omega}B^+_{\omega lm}\label{eq:Basym}
\end{align}
Notably, if we are in the regime where $\omega>0$, but $\tilde{\omega}<0$, then the reflection coefficients are greater than one. This is the superradiant sector of the solutions. Note that $m<0$ counter-rotating waves do not exhibit superradiance.

Because we are interested in two-point functions, we need to quantize the scalar field. In order to canonically quantize a scalar field on a curved background, we need a complete set of positive frequency solutions which are orthonormal with respect to the Klein-Gordon norm. The normalization is dominated by the asymptotic value of the solutions, and we find that for $\mathcal{R}=\mathcal{R}^+$,
\begin{align}\label{eq:in}
\Phi^{\text{in}}_{\omega lm}&=\frac{1}{\sqrt{8\pi^2\omega}}e^{-i\omega t}e^{im\phi}\mathcal{S}_{\omega lm}(\theta)\frac{\mathcal{R}^+_{\omega lm}(r)}{\sqrt{r^2+a^2}}\\
&(\Phi^{\text{in}}_{\omega lm},\Phi^{\text{in}}_{\omega' l'm'})=\delta(\omega-\omega')\delta_{ll'}\delta_{mm'}\quad\quad\omega>0
\end{align}
On the other hand, the set of solutions for which $\mathcal{R}=\mathcal{R}^-$ run into a subtle issue: waves in the superradiant regime have an overall negative norm. In particular, a superradiant mode can have a positive frequency with respect to an observer out at future null infinity but a negative frequency with respect to an observer on the past horizon. This is in stark contrast to fields in Schwarzschild, where positive frequency with respect to an asymptotic observer implies automatically implies a positive frequency at the horizon. For this reason, it is often common practice to define vacuum states in Schwarzschild by their properties on past and future null infinity or on the past and future horizons. The takeaway from Kerr is that we should only define properties of states along a Cauchy surface. The way to resolve the negative norm is to treat the superradiant sector separately from the other modes, namely
\begin{alignat}{3}
\Phi^{\text{up}}_{\omega lm}&=\frac{1}{\sqrt{8\pi^2\tilde{\omega}}}e^{-i\omega t}e^{im\phi}\mathcal{S}_{\omega lm}(\theta)\frac{\mathcal{R}^-_{\omega lm}(r)}{\sqrt{r^2+a^2}}\quad\quad\quad\quad&&\tilde{\omega}>0&&\\
&(\Phi^{\text{up}}_{\omega lm},\Phi^{\text{up}}_{\omega' l'm'})=\delta(\omega-\omega')\delta_{ll'}\delta_{mm'}\\
\Phi^{\text{up}}_{-\omega l-m}&=\frac{1}{\sqrt{8\pi^2(-\tilde{\omega})}}e^{i\omega t}e^{-im\phi}\mathcal{S}_{\omega lm}(\theta)\frac{\mathcal{R}^-_{-\omega l-m}(r)}{\sqrt{r^2+a^2}}&&0>\tilde{\omega}>-m\Omega&&\\
&(\Phi^{\text{up}}_{-\omega l-m},\Phi^{\text{up}}_{-\omega' l'-m'})=\delta(\omega-\omega')\delta_{ll'}\delta_{mm'}\label{eq:up}
\end{alignat} 
These define a set of orthonormal modes with positive norm. We call these "in" and "up" modes corresponding to their behavior according to \eqref{eq:Kasym}. The in modes correspond to a flux incoming from past null infinity, while the up modes correspond to a flux coming from the past horizon. There also exists another useful set of orthonormal modes. $V_{\omega lm}$ in \eqref{eq:Krad} is real, so $\mathcal{R}_{\omega lm}^*(r)$  also constitutes a solution to the radial equation. This new set of ``out" and ``down" modes have the same structure as the in and up modes in \Crefrange{eq:in}{eq:up} respectively, albeit replacing the radial function with its complex conjugate.

We are now ready to expand the scalar field in terms of its modes
\begin{align}
\Phi(x)=&\sum_{l,m}\left(\int_0^\infty d\omega(a^{in}_{\omega lm}\Phi^{in}_{\omega lm}+a^{in\dag}_{\omega lm}\Phi^{in\ast}_{\omega lm})+\int_{\omega_{min}}^\infty d\omega(a^{up}_{\omega lm}\Phi^{up}_{\omega lm}+a^{up\dag}_{\omega lm}\Phi^{up\ast}_{\omega lm})\right)\\
&+\sum_{l,m}\int_0^{\omega_{min}}d\omega(a^{up}_{-\omega l-m}\Phi^{up}_{-\omega l-m}+a^{up\dag}_{-\omega l-m}\Phi^{up\ast}_{-\omega l-m})
\end{align}
with $\omega_{min}=\text{max}\{0,m\Omega\}$. We quantize the field by promoting the coefficients to creation and annihilation operators which satisfy the typicaly commutation relations
\begin{equation}
\left[\hat{a}^{in}_{\omega lm},\hat{a}^{in\dag}_{\omega' l'm'}\right]=
\left[\hat{a}^{up}_{\omega lm},\hat{a}^{up\dag}_{\omega' l'm'}\right]=
\left[\hat{a}^{up}_{-\omega l-m},\hat{a}^{up\dag}_{-\omega' l'-m'}\right]=\delta(\omega-\omega')\delta_{ll'}\delta_{mm'}
\end{equation}
The state $\ket{B^-}$ which is annihilated by $\hat{a}^{in}_{\omega lm},\ \hat{a}^{up}_{\omega lm},\text{ and }\hat{a}^{up}_{-\omega l-m}$ is empty on the past horizon and past null infinity is called the ``past Boulware vacuum". In Schwarzschild, we ordinarily define the Boulware vacuum as the state which is empty on past and future null infinity, however, as explained above, it really only makes sense to consider states on Cauchy hypersurfaces. We could have also chosen to expand $\Phi$ in terms of the out and down modes. The vacuum state with respect to these modes is empty on the future horizon and future null infinity, and is aptly called the ``future Boulware vacuum". It is important to note that the in/up modes and the out/down modes are not orthogonal, and there is a simple Bogoliubov transformation between them. For this reason it is impossible to write down a vacuum state in Kerr which is empty at both past and future null infinity.

The next two states that we are interested in are the Unruh and Hartle-Hawking states. The Unruh vacuum is characterized as being empty at past null infinity, but thermal with respect to Rindler observers at the horizon. The Hartle-Hawking state is a Hadamard state which is regular everywhere and respects the symmetries of Kerr. In order to realize these properties, we need a new set of orthonormal modes in which we can expand the field operator. The modes were found by Frolov and Thorne \cite{Frolov_1989}, and it was shown that the two point functions in the Unruh and ``Frolov-Thorne" Hartle-Hawking state are
\begin{align}
G_{U}(x,x')&=\sum_{l,m}\left(\int d\tilde{\omega}\coth\left(\frac{2\tilde{\omega}}{T_H}\right)\Phi^{up}_{\omega lm}(x)\Phi^{up\ast}_{\omega lm}(x')+\int d\omega \Phi^{in}_{\omega lm}(x)\Phi^{in\ast}_{\omega lm}(x')\right)\\
G_{FT}(x,x')&=\sum_{l,m}\left(\int d\tilde{\omega}\coth\left(\frac{2\tilde{\omega}}{T_H}\right)\Phi^{up}_{\omega lm}(x)\Phi^{up\ast}_{\omega lm}(x')+\int d\omega \coth\left(\frac{2\tilde{\omega}}{T_H}\right)\Phi^{in}_{\omega lm}(x)\Phi^{in\ast}_{\omega lm}(x')\right)\notag
\end{align}
The Frolov-Thorne state is constructed so that it has the appropriate symmetries, and Frolov and Thorne claimed that the state is regular out to the speed-of-light surface. However, it was shown in \cite{Ottewill_2000} that although the expectation value of the (renormalized) stress tensor in the Frolov-Thorne state is regular on both the future and past null horizons, it fails to be regular off of the horizon in general. For example, the expectation value\footnote{We take the difference in the expectation value in order to subtract out state-independent divergent terms which would require renormalization.} of $\Phi^2$ for various states are
\begin{align}
\bra{B^-}\Phi(x)^2\ket{B^-}-\bra{U}\Phi(x)^2\ket{U}\underset{r\rightarrow\infty}{\sim}\sum_{l,m}\int d\tilde{\omega}\frac{-2}{\tilde{\omega}(e^{\tilde{\omega}/T_H}-1)}|B^-_{\omega lm}|^2|S_{\omega lm}(\theta)|^2 \label{eq:UBint}\\
\bra{U}\Phi(x)^2\ket{U}-\bra{FT}\Phi(x)^2\ket{FT}\underset{r\rightarrow r_+}{\sim}\sum_{l,m}\int d\omega\frac{-2}{\omega(e^{\tilde{\omega}/T_H}-1)}|B^+_{\omega lm}|^2|S_{\omega lm}(\theta)|^2 \label{eq:UFTint}
\end{align}
With Equation \eqref{eq:Basym} we can see that while the integrand of \eqref{eq:UBint} is regular, the integral over $\omega$ in \eqref{eq:UFTint} is divergent due to the pole at $\tilde{\omega}=0$. This implies that $\ket{FT}$ is not regular. The only exception to this is when at least one of the points in the correlation function is on the axis of rotation $\theta=0$. All modes with $m\neq 0$ vanish in this limit, and the above integral will no longer be divergent as $\tilde{\omega}\rightarrow\omega$. Notably, all of the superradiant modes drop out in this limit, and there is no longer an ambiguity concerning positive frequency and quantization.

\renewcommand{\baselinestretch}{1}\small
\bibliographystyle{ourbst}
\bibliography{NHEK_Bib}

\providecommand{\href}[2]{#2}\begingroup\raggedright\begin{thebibliography}{10}

\bibitem{Hartman_2020}
A.~Almheiri, T.~Hartman, J.~Maldacena, E.~Shaghoulian and A.~Tajdini, {{The
  entropy of Hawking radiation}},
  \href{http://dx.doi.org/10.1103/RevModPhys.93.035002}{Rev. Mod. Phys. {\bf
  93}, 035002, 2021},
  [\href{http://arxiv.org/abs/arXiv:2006.06872}{{arXiv:2006.06872 [hep-th]}}].

\bibitem{Bousso_2022}
R.~Bousso, X.~Dong, N.~Engelhardt, T.~Faulkner, T.~Hartman, S.~H. Shenker and
  D.~Stanford, {Snowmass white paper: Quantum aspects of black holes and the
  emergence of spacetime},  2022.

\bibitem{Faulkner_2016}
T.~Faulkner, R.~G. Leigh, O.~Parrikar and H.~Wang, {Modular hamiltonians for
  deformed half-spaces and the averaged null energy condition},
  \href{http://dx.doi.org/10.1007/jhep09(2016)038}{{\bf 2016}, Journal of High
  Energy Physics, 2016}.

\bibitem{Hartman_2017}
T.~Hartman, S.~Kundu and A.~Tajdini, {Averaged null energy condition from
  causality}, \href{http://dx.doi.org/10.1007/jhep07(2017)066}{{\bf 2017},
  Journal of High Energy Physics, 2017}.

\bibitem{Graham_2007}
N.~Graham and K.~D. Olum, {Achronal averaged null energy condition},
  \href{http://dx.doi.org/10.1103/physrevd.76.064001}{{\bf 76}, Physical Review
  D, 2007}.

\bibitem{Gao_2017}
P.~Gao, D.~L. Jafferis and A.~C. Wall, {Traversable wormholes via a double
  trace deformation}, \href{http://dx.doi.org/10.1007/jhep12(2017)151}{{\bf
  2017}, Journal of High Energy Physics, 2017}.

\bibitem{Maldacena_2017}
J.~Maldacena, D.~Stanford and Z.~Yang, {Diving into traversable wormholes},
  \href{http://dx.doi.org/10.1002/prop.201700034}{Fortschritte der Physik {\bf
  65}, 1700034, 2017}.

\bibitem{maldacena2018eternal}
J.~Maldacena and X.-L. Qi, {{Eternal traversable wormhole}},  2018,
  [\href{http://arxiv.org/abs/arXiv:1804.00491}{{arXiv:1804.00491 [hep-th]}}].

\bibitem{Caceres_2018}
E.~Caceres, A.~S. Misobuchi and M.-L. Xiao, {Rotating traversable wormholes in
  {AdS}}, \href{http://dx.doi.org/10.1007/jhep12(2018)005}{{\bf 2018}, Journal
  of High Energy Physics, 2018}.

\bibitem{Bak_2018}
D.~Bak, C.~Kim and S.-H. Yi, {Bulk view of teleportation and traversable
  wormholes}, \href{http://dx.doi.org/10.1007/jhep08(2018)140}{{\bf 2018},
  Journal of High Energy Physics, 2018}.

\bibitem{Ahn_2022}
B.~Ahn, S.-E. Bak, V.~Jahnke and K.-Y. Kim, {{Holographic teleportation with
  conservation laws: diffusion on traversable wormholes}},  2022,
  [\href{http://arxiv.org/abs/arXiv:2206.03434}{{arXiv:2206.03434 [hep-th]}}].

\bibitem{Al_Balushi_2021}
A.~A. Balushi, Z.~Wang and D.~Marolf, {Traversability of multi-boundary
  wormholes}, \href{http://dx.doi.org/10.1007/jhep04(2021)083}{{\bf 2021},
  Journal of High Energy Physics, 2021}.

\bibitem{Emparan_2021}
R.~Emparan, B.~Grado-White, D.~Marolf and M.~Toma{\v{s}}evi{\'{c}},
  {Multi-mouth traversable wormholes},
  \href{http://dx.doi.org/10.1007/jhep05(2021)032}{{\bf 2021}, Journal of High
  Energy Physics, 2021}.

\bibitem{Horowitz_2019}
G.~T. Horowitz, D.~Marolf, J.~E. Santos and D.~Wang, {Creating a traversable
  wormhole}, \href{http://dx.doi.org/10.1088/1361-6382/ab436f}{Classical and
  Quantum Gravity {\bf 36}, 205011, 2019}.

\bibitem{Fu_2019}
Z.~Fu, B.~Grado-White and D.~Marolf, {A perturbative perspective on
  self-supporting wormholes},
  \href{http://dx.doi.org/10.1088/1361-6382/aafcea}{Classical and Quantum
  Gravity {\bf 36}, 045006, 2019}.

\bibitem{McBride_2019}
D.~Marolf and S.~McBride, {Simple perturbatively traversable wormholes from
  bulk fermions}, \href{http://dx.doi.org/10.1007/jhep11(2019)037}{{\bf 2019},
  Journal of High Energy Physics, 2019}.

\bibitem{Marolf_2019}
Z.~Fu, B.~Grado-White and D.~Marolf, {Traversable asymptotically flat wormholes
  with short transit times},
  \href{http://dx.doi.org/10.1088/1361-6382/ab56e4}{Classical and Quantum
  Gravity {\bf 36}, 245018, 2019}.

\bibitem{Jafferis_2022}
D.~Jafferis, A.~Zlokapa, J.~D. Lykken, D.~K. Kolchmeyer, S.~I. Davis, N.~Lauk,
  H.~Neven and M.~Spiropulu, {Traversable wormhole dynamics on a quantum
  processor}, \href{http://dx.doi.org/10.1038/s41586-022-05424-3}{Nature {\bf
  612}, 51--55, 2022}.

\bibitem{maldacena2018traversable}
J.~Maldacena, A.~Milekhin and F.~Popov, {{Traversable wormholes in four
  dimensions}},  2018,
  [\href{http://arxiv.org/abs/arXiv:1807.04726}{{arXiv:1807.04726 [hep-th]}}].

\bibitem{maldacena2020humanly}
J.~Maldacena and A.~Milekhin, {Humanly traversable wormholes},
  \href{http://dx.doi.org/10.1103/physrevd.103.066007}{{\bf 103}, Physical
  Review D, 2021}.

\bibitem{maldacena2020comments}
J.~Maldacena, {Comments on magnetic black holes},
  \href{http://dx.doi.org/10.1007/jhep04(2021)079}{{\bf 2021}, Journal of High
  Energy Physics, 2021}.

\bibitem{Horowitz_2023}
G.~T. Horowitz, M.~Kolanowski, G.~N. Remmen and J.~E. Santos, {Extremal kerr
  black holes as amplifiers of new physics},  2023.

\bibitem{Ottewill_2000}
A.~C. Ottewill and E.~Winstanley, {Renormalized stress tensor in kerr
  space-time: General results},
  \href{http://dx.doi.org/10.1103/physrevd.62.084018}{{\bf 62}, Physical Review
  D, 2000}.

\bibitem{Bardeen_1999}
J.~Bardeen and G.~T. Horowitz, {Extreme {Kerr} throat geometry: A vacuum analog
  of $\text{AdS}_2\times \text{S}^2$},
  \href{http://dx.doi.org/10.1103/physrevd.60.104030}{{\bf 60}, Physical Review
  D, 1999}.

\bibitem{Bredberg_2010}
I.~Bredberg, T.~Hartman, W.~Song and A.~Strominger, {Black hole superradiance
  from {Kerr/CFT}}, \href{http://dx.doi.org/10.1007/jhep04(2010)019}{{\bf
  2010}, Journal of High Energy Physics, 2010}.

\bibitem{Anninos_2009}
D.~Anninos, W.~Li, M.~Padi, W.~Song and A.~Strominger, {Warped $\text{AdS}_3$
  black holes}, \href{http://dx.doi.org/10.1088/1126-6708/2009/03/130}{Journal
  of High Energy Physics {\bf 2009}, 130--130, 2009}.

\bibitem{Guica_2009}
M.~Guica, T.~Hartman, W.~Song and A.~Strominger, {The {Kerr/CFT}
  correspondence}, \href{http://dx.doi.org/10.1103/physrevd.80.124008}{{\bf
  80}, Physical Review D, 2009}.

\bibitem{Rasmussen_2010}
J.~Rasmussen, {A near-{NHEK/CFT} correspondence},
  \href{http://dx.doi.org/10.1142/s0217751x10051001}{International Journal of
  Modern Physics A {\bf 25}, 5517--5527, 2010}.

\bibitem{Hartman_2009}
T.~Hartman, W.~Song and A.~Strominger, {{Holographic Derivation of Kerr-Newman
  Scattering Amplitudes for General Charge and Spin}},
  \href{http://dx.doi.org/10.1007/JHEP03(2010)118}{JHEP {\bf 03}, 118, 2010},
  [\href{http://arxiv.org/abs/arXiv:0908.3909}{{arXiv:0908.3909 [hep-th]}}].

\bibitem{Aharony_2005}
O.~Aharony, M.~Berkooz and B.~Katz, {Non-local effects of multi-trace
  deformations in the {AdS/CFT} correspondence},
  \href{http://dx.doi.org/10.1088/1126-6708/2005/10/097}{Journal of High Energy
  Physics {\bf 2005}, 097, 2005}.

\bibitem{Strominger_1999}
J.~Maldacena, J.~Michelson and A.~Strominger, {Anti-de sitter fragmentation},
  \href{http://dx.doi.org/10.1088/1126-6708/1999/02/011}{Journal of High Energy
  Physics {\bf 1999}, 011--011, 1999}.

\bibitem{Lucietti_2012}
J.~Lucietti and H.~S. Reall, {Gravitational instability of an extreme kerr
  black hole}, \href{http://dx.doi.org/10.1103/physrevd.86.104030}{{\bf 86},
  Physical Review D, 2012}.

\bibitem{Aretakis_2013}
S.~Aretakis, {A note on instabilities of extremal black holes under scalar
  perturbations from afar},
  \href{http://dx.doi.org/10.1088/0264-9381/30/9/095010}{Classical and Quantum
  Gravity {\bf 30}, 095010, 2013}.

\bibitem{Wald_1988}
B.~S. Kay and R.~M. Wald, {{Theorems on the Uniqueness and Thermal Properties
  of Stationary, Nonsingular, Quasifree States on Space-Times with a Bifurcate
  Killing Horizon}},
  \href{http://dx.doi.org/10.1016/0370-1573(91)90015-E}{Phys. Rept. {\bf 207},
  49--136, 1991}.

\bibitem{Candelas_1981}
P.~Candelas, P.~Chrzanowski and K.~W. Howard, {{Quantization of Electromagnetic
  and Gravitational Perturbations of a Kerr Black Hole}},
  \href{http://dx.doi.org/10.1103/PhysRevD.24.297}{Phys. Rev. D {\bf 24},
  297--304, 1981}.

\bibitem{Frolov_1989}
V.~P. Frolov and K.~S. Thorne, {{Renormalized Stress - Energy Tensor Near the
  Horizon of a Slowly Evolving, Rotating Black Hole}},
  \href{http://dx.doi.org/10.1103/PhysRevD.39.2125}{Phys. Rev. D {\bf 39},
  2125--2154, 1989}.

\bibitem{Winstanley_2023}
V.~Balakumar, R.~Bernar and E.~Winstanley, {Superradiance and quantum states on
  black hole space-times},  2023.

\bibitem{Winstanley_2013}
M.~Casals, S.~R. Dolan, B.~C. Nolan, A.~C. Ottewill and E.~Winstanley,
  {Quantization of fermions on {Kerr} space-time},
  \href{http://dx.doi.org/10.1103/physrevd.87.064027}{{\bf 87}, Physical Review
  D, 2013}.

\bibitem{Gerard_2021}
C.~Gérard, D.~Häfner and M.~Wrochna, {The unruh state for massless fermions
  on kerr spacetime and its hadamard property},  2021.

\bibitem{Dias_2009}
O.~J. Dias, H.~S. Reall and J.~E. Santos, {{Kerr-CFT} and gravitational
  perturbations},
  \href{http://dx.doi.org/10.1088/1126-6708/2009/08/101}{Journal of High Energy
  Physics {\bf 2009}, 101, 2009}.

\bibitem{Amsel_2009}
A.~J. Amsel, G.~T. Horowitz, D.~Marolf and M.~M. Roberts, {No dynamics in the
  extremal kerr throat},
  \href{http://dx.doi.org/10.1088/1126-6708/2009/09/044}{Journal of High Energy
  Physics {\bf 2009}, 044, 2009}.

\bibitem{NAKATSU_1999}
T.~Nakatsu and N.~Yokoi, {Comments on hamiltonian formalism of {AdS/CFT}
  correspondence}, \href{http://dx.doi.org/10.1142/s0217732399000183}{Modern
  Physics Letters A {\bf 14}, 147--159, 1999}.

\bibitem{Spradlin_1999}
M.~Spradlin and A.~Strominger, {Vacuum states for ${AdS}_2$ black holes},
  \href{http://dx.doi.org/10.1088/1126-6708/1999/11/021}{Journal of High Energy
  Physics {\bf 1999}, 021, 1999}.

\bibitem{Gralla_2018}
S.~E. Gralla, A.~Lupsasca and A.~Strominger, {Observational signature of high
  spin at the event horizon telescope},
  \href{http://dx.doi.org/10.1093/mnras/sty039}{Monthly Notices of the Royal
  Astronomical Society {\bf 475}, 3829--3853, 2018}.

\bibitem{Kapec_2019}
D.~Kapec and A.~Lupsasca, {Particle motion near high-spin black holes},
  \href{http://dx.doi.org/10.1088/1361-6382/ab519e}{Classical and Quantum
  Gravity {\bf 37}, 015006, 2019}.

\bibitem{Lupsasca_2020}
A.~Lupsasca, D.~Kapec, Y.~Shi, D.~E.~A. Gates and A.~Strominger, {Polarization
  whorls from $m87^{\ast}$ at the event horizon telescope},
  \href{http://dx.doi.org/10.1098/rspa.2019.0618}{Proceedings of the Royal
  Society A: Mathematical, Physical and Engineering Sciences {\bf 476},
  20190618, 2020}.

\bibitem{Hadar_2022}
S.~Hadar, D.~Kapec, A.~Lupsasca and A.~Strominger, {Holography of the photon
  ring}, \href{http://dx.doi.org/10.1088/1361-6382/ac8d43}{Classical and
  Quantum Gravity {\bf 39}, 215001, 2022}.

\bibitem{DeWitt_1975}
B.~S. DeWitt, {{Quantum Field Theory in Curved Space-Time}},
  \href{http://dx.doi.org/10.1016/0370-1573(75)90051-4}{Phys. Rept. {\bf 19},
  295--357, 1975}.

\end{thebibliography}\endgroup
\end{document}